# Periodically Aligned Liquid Crystal:

# Potential application for projection displays

H. Sarkissian[1], N. Tabirian[2], B. Park[1], and B. Zeldovich[1,2].

[1] College of Optics / CREOL, University of Central Florida,

4000 Central Florida Blvd., Orlando, FL 32816-2700,

and

[2] BEAM Corp.,686 Formosa Ave., Winter Park, FL 32789

boris@creol.ucf.edu, hakob@creol.ucf.edu

voice: (407) 823-6831, (407) 823-6806; FAX (407) 823-6880

Keywords: Liquid crystals, Diffractive displays, Circular Polarization.

Abstract:

A nematic liquid crystal (NLC) layer with the anisotropy axis modulated at a fixed rate $q$ in the transverse direction is considered. If the layer locally constitutes a half-wave plate, then the thin-screen approximation predicts 100% -efficient diffraction of normal incident wave. The possibility of implementing such a layer via anchoring at both surfaces of a cell with thickness $L$ is studied as a function of parameter $qL$ and threshold values of this parameter are found for a variety of cases. Distortions of the structure of director in comparison with the preferable ideal profile are found via numerical modeling. Freedericksz transition is studied for this configuration. Coupled-mode theory is applied to light propagation through such cell allowing to account for walk-off effects and effects of nematic distortion. In summary, this cell is suggested as a means for projection display; high efficiency is predicted.

## 1. Introduction

Technology of projection displays exhibited enormous progress in the past decade. In particular, Liquid Crystal-based projection Displays (projection LCDs), [1, 2], actively compete with the Micro-Electro-Mechanical Systems (MEMS) [3], and with older technologies like projection from the screens of Cathode Ray Tubes (CRTs), [4].

An interesting subset of the design of the projection LCDs is connected with the following idea. In the "off" state of a pixel, the illuminating beam is diffracted at a considerable angle. As a result the beam is blocked by an aperture positioned downstream of the light propagation, [5-9]. The pixel can be switched into the "on" state by eliminating the diffraction using a controlling agent; typically a locally applied electric field. One of the problems in the above design of the projection displays is the requirement of very strong suppression of the intensity left in the zeroth diffraction order in the "off" state, so that high contrast ratio may be achieved.

In this paper we propose the use of an LC cell with a specific orientation of director. Namely, we suggest using a cell, with its director's azimuth monotonously rotating as a function of one of the transverse coordinates. When the cell locally constitutes a "half-wave plate" with respect to polarization, the diffraction of incident light from the incident diffraction order is complete. Additional advantage of this design of the element of a projection LCD is that it <u>utilizes both polarizations</u> of incident light.

In Section 2 of this paper we describe the physical principle of the operation of the suggested NLC element. We develop a detailed study of the performance of the suggested display element in the rest of the paper.

In Section 3 we study distortions of the NLC orientation, i.e. deviations of the director profile in the bulk of NLC cell from its ideal structure given by Eq. 1. We show that the preferred orientation is stable, if the ratio of thickness to the modulation period ($L/\Lambda$) is small enough, typically less or about 0.4. We also consider the change of this profile under the application of the controlling electric field (Friedericksz transition).

These results are obtained, using numeric integration of the dynamic equations to find the equilibrium director profile of NLC in the cell.

Then, in section 4 we consider the problem of light propagation in the cell with the account of the walk-off effects. The latter effects arise due to the finite angular tilt of the incident and diffracted waves in the process of propagation, so that the tilted ray traverses considerable fraction of the structure's period. We estimate the impact of the walk-off effects on performance of the cell. In particular, we elucidate the potentially harmful influence of those effects on the degree of suppression of transmitted light in the "off" state, i.e. on the contrast ratio. Finally, we make a similar estimation for the influence of the bulk distortions of the NLC director profile, found in section 3, on diffraction.

## 2. Principle of operation of projection display element

Fig. 1a, b. We consider nematic liquid crystal, confined inside a layer $0 < z < L$, with the azimuth of the director $\mathbf{n}$ monotonously rotating in transverse direction $x$, Figure 1, so that a periodic structure

$$\mathbf{n} = \mathbf{n}_0(x) = \{\cos(qx), \sin(qx), 0\}, \qquad q = 2\pi/\Lambda, \tag{1}$$

is realized. Here $\Lambda$ is the director's period. Since NLC's vector of director $\mathbf{n}$ and its negative, $-\mathbf{n}$, are physically equivalent, physical properties of such NLC are modulated with the period $\Lambda/2$. Such NLC is locally birefringent, with the optical axis lying in $(x, y)$-plane and rotating along the $x$ direction. Let's consider the effect of this birefringence on normally incident light using the Jones matrix analysis.

According to Jones matrix formulation, the evolution of electric field $\mathbf{E}(z)$ along the propagation direction is given by

$$\begin{pmatrix} E_x(z) \\ E_y(z) \end{pmatrix} = \hat{T}(x, z) \cdot \begin{pmatrix} E_x(0) \\ E_y(0) \end{pmatrix}.$$

(2)

Here $\mathbf{E}(z)$ is slowly varying envelope of the electric field at the optical frequency $\omega$,

$$\mathbf{E}_{real}(z, t) = 0.5[\mathbf{E}(z)\exp(-i\omega t + ikz) + \text{compl. conj.}].$$

Matrix **T** represents phase retardation by the birefringent medium with refractive indexes $n_o$ and $n_e$, respectively. It should be written in local coordinate system attached to the director, so that

$$\hat{T}(x,z) = \begin{pmatrix} \cos qx & -\sin qx \\ \sin qx & \cos qx \end{pmatrix} \cdot \begin{pmatrix} \exp(i\kappa \cdot z) & 0 \\ 0 & \exp(-i\kappa \cdot z) \end{pmatrix} \cdot \begin{pmatrix} \cos qx & \sin qx \\ -\sin qx & \cos qx \end{pmatrix}, \quad \kappa = \pi \cdot (n_e - n_o)/\lambda. \quad (3)$$

Here $2\kappa \cdot z$ is the phase retardation between the two polarizations, and $\lambda$ is the wavelength in vacuum. Thus the evolution of the electric field with distance is given by

$$E_x(x, z) = [\cos(z \cdot \kappa) + i\cos(2qx)\sin(z \cdot \kappa)]E_x(z=0) + i\sin(2qx)\sin(z \cdot \kappa)E_y(z=0),$$

$$E_y(x, z) = i\sin(2qx)\sin(z \cdot \kappa)E_x(z=0) + [\cos(z \cdot \kappa) - i\cos(2qx)\sin(z \cdot \kappa)]E_y(z=0).$$

These relations can be written in terms of circularly-polarized components:

$$R = E_x - iE_y, \qquad L = E_x + iE_y, \quad (4)$$

$$L(x, z) = \cos(z \cdot \kappa) L(z=0) + i \cdot \exp(2iqx) \sin(z \cdot \kappa) R(z=0)$$

$$R(x, z) = \cos(z \cdot \kappa) R(z=0) + i \cdot \exp(-2iqx) \sin(z \cdot \kappa) L(z=0)$$

If we assume

$$L(x, z) = L_0(z) + L_1(z)\exp(2iqx) \text{ and } R(x, z) = R_0(z) + R_{-1}(z)\exp(-2iqx), \quad (5)$$

then

$$R_0(z) = \cos(z \cdot \kappa) R_0(0) + i \sin(z \cdot \kappa) L_1(0); \qquad L_1(z) = \cos(z \cdot \kappa) L_1(0) + i \sin(z \cdot \kappa) R_0(0). \quad (6)$$

$$R_{-1}(z) = \cos(z \cdot \kappa) R_{-1}(0) + i \sin(z \cdot \kappa) L_0(0); \qquad L_0(z) = \cos(z \cdot \kappa) L_0(0) + i \sin(z \cdot \kappa) R_{-1}(0), \quad (7)$$

Relationships (6, 7) describe the process of coupling between two circularly-polarized waves, $R$ and $L$, propagating in the modulated NLC. It shows that only the plus first and minus first diffraction orders are generated by the device. If the thickness of the cell $L$ satisfies the condition of half-wave plate,

$$L(n_e - n_o) = \lambda/2, \text{ i.e. } \kappa \cdot L = \pi/2,$$

then all power of incident wave with right circular polarization is transferred into the (+1$^{st}$) diffraction order, and that diffraction order acquires left circular polarization. Similarly, incident wave with left circular polarization is 100% transferred into the ($-1^{st}$) diffraction order, and that diffraction order acquires right circular polarization,

$$R_{-1}(z=L) = iL_0(z=0), \qquad L_1(z=L) = iR_0(z=0). \tag{8}$$

This important property of the sinusoidally aligned NLC has been previously discussed [10-12]. It constitutes the basis of the use of the NLC structure (1) as an element of a projection display. In configuration shown on Figure 2 we propose a possible design of such display cell. Transparent electrodes are attached to the cell surfaces. When no voltage is applied and the NLC structure is periodically aligned, the waves do diffract and can be blocked with an aperture. In this case the cell works in the "off" regime. The cell can be switched to the "on" regime by application of voltage sufficiently high to destroy the periodic structure. Then the NLC cell becomes homogeneous, light is not diffracted, and passes through the aperture.

In addition to this application, this element can be used as an electrically controllable polarization beam splitter since the only diffracted orders are the two first orders, which are right- and left-circularly polarized. If, however, the cell is illuminated by two right- and left-circularly polarized waves, corresponding to plus- and minus-first diffracted orders, they combine into a single beam. In this way, this cell may be used as an electrically controllable incoherent beam combiner.

In the rest of the paper we develop detailed study of the performance of the suggested element.

## 3. Equilibrium of the director in the periodically aligned NLC cell

Possible way to obtain the desired NLC structure as in Eq. (1) is through anchoring at the top and the bottom interfaces. Prescribed angle of anchoring of NLC at the boundary is usually achieved by mechanical rubbing of the surface coated with a very thin polymer layer. Orientation of Eq (1) evidently cannot be realized with rubbing. However, there is another method to create the prescribed anchoring azimuth which utilizes photo-polymerization under illumination by a linearly-polarized UV light, [13-16]. Monotonously rotating azimuth of local linear polarization of the UV light may be achieved, if light constitutes a pair of two mutually coherent circularly polarized waves, one right circular, and the other left circular, propagating at an angle to each other. Indeed,

$$\mathbf{E}(x) = E_0(\mathbf{e_x} + i\mathbf{e_y})\exp(-iqx) + E_0(\mathbf{e_x} - i\mathbf{e_y})\exp(iqx) \equiv 2E_0[\mathbf{e_x}\cos(qx) + \mathbf{e_y}\sin(qx)] = \text{const}\cdot\mathbf{n}_0(x). \quad (9)$$

The rest of this Section 3 deals with the problem of stability of the desired structure of the director, Eq. (1), in the bulk of the NLC cell. When thickness $L$ is small, the director distribution in the bulk should follow the distribution prescribed at the surfaces. However, such distribution is unstable for a thick cell, since its elastic energy can be reduced by formation of a homogeneous distribution in the bulk. The onset of this instability is characterized by the value of critical thickness, over which structure Eq. (1) becomes impossible to achieve.

In order to obtain the director distribution in the NLC which is anchored at the boundaries and to analyze its stability, we have to solve the equilibrium equations with appropriate boundary conditions. This solution is convenient to achieve via consideration of time-dependent process of build-up of the steady-state distribution of the director.

Dynamics of the director of the NLC can be described with the evolution equation, [17, 18],

$$\frac{\delta F}{\delta n_i} - \partial_k \frac{\delta F}{\delta(\partial_k n_i)} = \frac{\delta R}{\delta \dot{n}_i}, \quad (10)$$

Here $R$ is the dissipative function, and $F$ is the nematic free energy, respectively:

$$R = \gamma\{(\partial \mathbf{n}/\partial t)\cdot(\partial \mathbf{n}/\partial t)\}/2, \qquad F = (1/2)\cdot[K_1(\text{div }\mathbf{n})^2 + K_2(\mathbf{n}\cdot\text{curl }\mathbf{n})^2 + K_3(\mathbf{n}\times\text{curl }\mathbf{n})^2]. \quad (11)$$

After substituting (11) into (10) one can write the evolution equation in terms of director $\mathbf{n}$

$$\partial\mathbf{n}/\partial t = \mathbf{h}/\gamma, \qquad \mathbf{h} = \mathbf{H} - \mathbf{n}(\mathbf{n}\cdot\mathbf{H}). \quad (12)$$

Vector $\mathbf{H}$ is known as the "molecular field" before the procedure of its orthogonalization to local director $\mathbf{n}$, while vector $\mathbf{h}$ is the "molecular field' after that procedure. Cartesian components of $\mathbf{H}$ are

$$H_i = (K_1 - K_2)\partial_i(\partial_\alpha n_\alpha) + K_2(\partial_\alpha\partial_\alpha)n_i + (K_3 - K_2)\{n_\alpha n_\beta(\partial_\alpha\partial_\beta n_i) + n_\alpha(\partial_\beta n_\beta)(\partial_\alpha n_i) +$$

$$+ n_\alpha(\partial_\alpha n_\beta)(\partial_\beta n_i) - n_\alpha(\partial_\alpha n_\beta)(\partial_i n_\beta)\} \quad (13)$$

The solution we are looking for should satisfy the boundary conditions at the surfaces $z = 0$ and $z = L$,

$$\mathbf{n}(x, z=0) = \mathbf{n}(x, z=L) = \mathbf{n}_0(x) \equiv \{\cos(qx), \sin(qx), 0\}, \qquad (14)$$

In the general case this equation, as written through the director and its spatial derivatives, cannot be solved analytically. However, an analytical solution exists in monoconstant case.

*Solution and its stability in monoconstant approximation*

In monoconstant approximation, when $K_1 = K_2 = K_3 = K$, the molecular field (13) is reduced to

$$\mathbf{H} = K(\nabla \cdot \nabla)\mathbf{n} \qquad (15)$$

Direct substitution shows that the director distribution $\mathbf{n}_0(x) = \{\cos(qx), \sin(qx), 0\}$ prescribed by the Eq. (1) satisfies the equilibrium condition in the whole volume of the cell, since the $\mathbf{H}$-vector turns out to be parallel to $\mathbf{n}_0(x)$, and hence $\mathbf{h}$-vector is zero:

$$\partial \mathbf{n}_0 / \partial t = (K/\gamma)\{(\nabla \cdot \nabla)\mathbf{n}_0 - \mathbf{n}_0[\mathbf{n}_0 \cdot (\nabla \cdot \nabla)\mathbf{n}_0]\} \equiv 0,$$

Therefore $\mathbf{n}_0(x)$ is an exact solution in the monoconstant case. The onset of instability of this solution can be evaluated by assuming

$$\mathbf{n}(x, z, t) = \{\cos(\beta)\cos(qx+\alpha), \cos(\beta)\sin(qx+\alpha), \sin(\beta)\}$$

where $\alpha(x, z, t)$ and $\theta(x, z, t)$ are small deviation angles. It can be approximated with

$$\mathbf{n}(x, z, t) \approx \mathbf{n}_0(x) + \alpha(x, z, t)\mathbf{m}(x) + \mathbf{e}_z\beta(x, z, t), \qquad \mathbf{m}(x) = \{-\sin(qx), \cos(qx), 0\} \qquad (16)$$

As a result one can obtain

$$\partial\beta/\partial t = (K/\gamma)[q^2\beta + (\nabla \cdot \nabla)\beta], \qquad \partial\alpha/\partial t = (K/\gamma)(\nabla \cdot \nabla)\alpha \qquad (17)$$

It is convenient to look for the solution of the $\beta$-equation in the form of a superposition of eigenmodes, $\exp(-\mu_m t)\sin(m\pi z/L)$. Then equation (17) yields

$$\beta \propto \sum\{\exp(-\mu_m t)\sin(m\pi z/L)\}, \qquad \mu_m = (K/\gamma)[(\pi m/L)^2 - q^2]. \qquad (18)$$

Therefore the onset of the instability begins (at the mode $m = 1$), when $L = 0.5\Lambda$, or $qL = \pi$. The system is always stable with respect to the $\alpha$-perturbations,.

*Numerical calculations.*

While monoconstant approximation is relatively simple and illustrative, it represents real NLCs only to some approximation. To examine real NLCs, we have solved the dynamic equations (12, 13) numerically. We assume arbitrary initial distribution with fixed periodic boundary conditions (14) at the surfaces. We then use Eq (12) to allow the LC to relax into the steady-state equilibrium distribution. When the ideal periodically aligned structure $\mathbf{n}_0(x)$ from Eq. (1) is stable, the volume average

$$\langle n_z \rangle = \frac{1}{V} \iiint_V n_z dV \tag{19}$$

is zero. It usually becomes non-zero in the presence of the considerable distortions in comparison with the ideal structure. Since the initial distribution and boundary conditions do not depend on the *y*-coordinate, we have omitted *y*-derivatives in our computation.

Fig. 3a, b.

Figures 3a and 4a demonstrate the plots of $\langle n_z \rangle$ as a function of thickness for monoconstant LC and for 5CB

Fig. 4a, b.

respectively. It has a pronounced critical point, which separates stable and unstable regions. The derivative of $\langle n_z \rangle$ with respect to parameter $(L/\Lambda)$ is presented in Figures 3b and 4b; this derivative has a sharp peak at the critical thickness. As expected, critical thickness is $L = 0.5\Lambda$ for monoconstant case. Table 1 lists values of critical thickness and maximum diffraction angle (calculated using Eq. (26) below) for a number of other NLC's.

Table 1.

*Asymptotic analytical solution: thin cell.*

When the thickness is well below critical, the sinusoidally aligned structure is reasonably close to the stable solution of the equations of equilibrium. The director may slightly deviate from the ideal unperturbed distribution due to the fact that the Frank's constants are not equal to each other. This deviation can be found using perturbation theory with small dimensionless parameter $qL$. Assuming $\mathbf{n} \approx \mathbf{n}_0 + \alpha \mathbf{m} + \mathbf{e}_z \theta$, as in Eq. (16), and keeping the terms up to $(qL)^2$ including, one can get

$$\mathbf{h}_0 + K_2 \mathbf{m} (\partial^2 \alpha / \partial z^2) = 0, \quad \mathbf{h}_0 = q^2 (K_1 - K_3) \sin(qx)\cos(qx) \cdot \mathbf{m}; \quad \beta \approx 0. \tag{20}$$

That allows to find the profile of azimuthal angle deviation α, if conditions α(at the boundaries) = 0 are assumed:

$$\alpha(x, z) = (qL)^2 \, 0.5 \, [(K_1 - K_3)/K_2] \sin(qx)\cos(qx) \, [z(L-z)/L^2] \qquad (21)$$

and therefore

$$\delta n_x(x, z) = -(qL)^2 \, 0.5 \, [(K_1 - K_3)/K_2] \sin^2(qx)\cos(qx) \, [z(L-z)/L^2] \qquad (22)$$

$$\delta n_y(x, z) = (qL)^2 \, 0.5 \, [(K_1 - K_3)/K_2] \sin(qx)\cos^2(qx) \, [z(L-z)/L^2] \qquad (23)$$

Numeric calculations confirm good validity of this perturbative solution for small $(qL)$, less or about 0.3. For larger values of that parameter, in the vicinity of the critical thickness, actual amplitude of perturbations was found to be much smaller than the perturbative analytic result. In particular, for the NLC 5CB and $qL \approx 2$, the amplitude of $n_x$, $n_y$-distortions was about 5 times smaller.

*Friedericksz transition in periodically bent NLC under the applied electric field.*

The pixel transmission is supposed to be controlled via application of external quasi-static (at radio-frequency) voltage $V$ across the layer $0 < z < L$, i.e. via the application of the field $\mathbf{E} = \mathbf{e_z}V/L$. Application of such field is accounted for by the addition of the term $-0.5\varepsilon_0(\varepsilon_\parallel - \varepsilon_\perp)(\mathbf{E}\cdot\mathbf{n})^2$ to the free energy density $F$ from Eq. (11). The analysis analogous to the one given above shows that the instability threshold is given by a line, which in monoconstant approximation is an ellipse in $(L, V)$-plane:

$$(qL)^2 + [\varepsilon_0 (\varepsilon_\parallel - \varepsilon_\perp)/K_2] V^2 = \pi^2, \qquad (L/L_0)^2 + (V/V_0)^2 = 1, \qquad (24)$$

where $L_0 = \Lambda/2$ and $V_0 = \{\pi^2 K_2 / [\varepsilon_0 (\varepsilon_\parallel - \varepsilon_\perp)]\}^{1/2}$. Here $\varepsilon_0$ is $8.85 \cdot 10^{-12}$ Farad/m, $\varepsilon_\parallel$ and $\varepsilon_\perp$ are the dimensionless longitudinal and transverse dielectric constants, respectively.

Fig. 5a, b. Figures 5a and 5b show the result of numerical build-up of the steady-state; namely the value of $\langle n_z \rangle$, as a function of cell thickness $L$ and applied voltage $V$ for a cell with monoconstant NLC and for 5CB respectively. On Figure 5a the stability area inside the ellipse is the region where the periodically aligned structure exists in ideal form Eq. (1). Our computations show that the form of such a region is close to ellipse, even if the NLC in consideration is not monoconstant.

## 4. Diffraction by periodically aligned NLC

As it was shown in section 2, achieving 100% diffraction efficiency requires that NLC layer locally constitutes λ/2 plate, i.e. that

$$L \cdot (n_e - n_o) = \lambda/2. \tag{25}$$

On the other hand, the diffraction angle can be found from the equation $\sin\theta = 2\lambda/\Lambda$, where taking half of the mathematical period $\Lambda$ accounts for the equivalence of plus and minus directions of the director. Thus,

$$\sin\theta = 2\lambda/\Lambda = 4(n_e - n_o)(L/\Lambda). \tag{26}$$

Through this equation the upper limit of $L/\Lambda$ imposes a limit on the maximum diffraction angle, shown in the Table 1.

The above conclusion about 100% diffraction efficiency was made in the approximation of a thin phase screen. In other words, up to now we have ignored the effects of walk-off of the ray across the periodic structure, as the light propagates at an angle ($\theta_{inc}/n$) to the z-direction inside the cell, see Figure 6. Another effect which may deteriorate the performance of the suggested device is the influence of possible deviations of the director from the ideal structure Eq. (1). To account for both of these effects, we have to write differential equations that describe 1) propagation of tilted waves, and 2) coupling of different polarizational / angular components through anisotropic dielectric permittivity tensor.

Fig. 6.

Since the values of the angle are relatively small, especially inside the material of the cell, we will use paraxial (parabolic) approximation for the dependence of z-component of wave vector $k_z$ on $\sin(\theta) \approx \theta$, [19]. We present the electric field of the light wave inside NLC in the form

$$\mathbf{E}_{real}(z, t) = 0.5[\mathbf{E}(x, z)\exp(-i\omega t + ikz) + \text{compl. conj.}], \quad k = (\omega n/c), \quad n = (n_e + n_o)/2, \tag{27}$$

$$\mathbf{E}(x, z) = 0.5(\mathbf{e_x} + i\mathbf{e_y})R(x, z) + 0.5(\mathbf{e_x} - i\mathbf{e_y})L(x, z), \tag{28}$$

$$R(x,z) = \exp(isx)\sum_{m=-\infty}^{\infty} R_m(z)\exp(2iqmx), \quad L(x,z) = \exp(isx)\sum_{m=-\infty}^{\infty} L_m(z)\exp(2iqmx), \tag{29}$$

where the amplitudes $R_m(x,z)$ and $L_m(x,z)$ satisfy equations

$$\frac{dR_m}{dz} = -\frac{i(s+2mq)^2}{2k} R_m(z) + i\sum_p R_{m+p}(z)\rho_p + i\sum_p L_{m+p}(z)\kappa_p$$

(30)

$$\frac{dL_m}{dz} = -\frac{i(s+2mq)^2}{2k} L_m(z) + i\sum_p L_{m+p}(z)\sigma_p + i\sum_p R_{m-p}(z)(\kappa_p)^* \tag{31}$$

The variable $s = (\omega/c)\theta_{inc}$ characterizes the angle of incidence of the input wave (in air). The terms $(s+2mq)^2/2k$ in Eqs. (30, 31) constitute the difference between $k_z$, $z$-projection of the wave vector for the given $m$-th Fourier component in the expansion (29) and the value $k = (\omega n/c)$, which was assumed in Eq. (27) for normal incidence. To be more precise, we took the paraxial (parabolic) approximation for this difference. These terms describe propagation of tilted beams and, thus, walk-off effects. Coupling coefficients $\rho_p$, $\sigma_p$ and $\kappa_p$ describe the influence of optical anisotropy of liquid crystal in periodically aligned configuration.

The approximation of ideal structure Eq. (1) corresponds to only one non-zero coupling coefficient:

$$\kappa_1 = (\kappa_1)^* = \kappa = \pi \cdot (n_e - n_o)/\lambda. \tag{32}$$

Meanwhile all the other coupling coefficients $\rho_p$, $\sigma_p$ and $\kappa_p$ are zero for propagation in the ideal structure Eq(1). Then the system (30, 31) for the incident $R_0$-wave simplifies to

$$dR_0(z)/dz = -i\nu_0 R_0(z) + i\,\kappa \cdot L_1(z); \qquad dL_1(z)/dz = i\,\kappa \cdot R_0(z) - i\nu_1 L_1(z), \tag{33}$$

$$\nu_0 = s^2/2k, \qquad \nu_1 = (s+2q)^2/2k, \tag{34}$$

Solution of the system (30), (31) is well-known:

$$\begin{pmatrix} R_0(z) \\ L_1(z) \end{pmatrix} = \hat{M}(z) \cdot \begin{pmatrix} R_0(z=0) \\ L_1(z=0) \end{pmatrix}, \quad \hat{M} = \exp(i\varphi)\begin{pmatrix} \cos(uz) - i(\xi/u)\sin(uz) & i\kappa\sin(uz)/u \\ i\kappa\sin(uz)/u & \cos(uz) + i(\xi/u)\sin(uz) \end{pmatrix} \tag{35}$$

$$\varphi = -(\nu_0 + \nu_1)/2, \qquad \xi = (\nu_0 - \nu_1)/2 = -(sq + q^2)/k, \qquad u = (\kappa^2 + \xi^2)^{0.5}, \tag{36}$$

Those expressions show that the diffraction efficiency $\eta$, i.e. intensity coefficient of transformation from $R_0$ into $L_1$,

$$\eta = [(\kappa/u) \sin(uz)]^2 \qquad (37)$$

has its maximum at $\xi = 0$ for any value of $z$. in the range $0 < z < \pi/2\kappa$. Moreover, 100% efficiency is achieved, when $\kappa L = \pi \cdot (n_e - n_o)/\lambda \, L = \pi/2$, i.e. when the condition Eq.(25) of half-wave plate is satisfied. In view of our task of creating an element for a display pixel, we are mostly concerned with leaving no intensity in the original, un-diffracted order. This remaining intensity will constitute the following fraction of the incident intensity $|R_0|^2$:

$$|R_0(z=L)|^2 = (1-\eta) \cdot |R_0(z=0)|^2 \approx |R_0(z=0)|^2 \cdot [(2\xi L/\pi)^2 + (\kappa L - \pi/2)^2]. \qquad (38)$$

We see that even for the optimum value of interaction strength, $\kappa L = \pi/2$, remaining transmission is about $0.4(\xi L)^2$. In its turn, the optimum value $\xi = 0$ is achieved at symmetric geometry, $\theta_{inc} = -(\lambda/\Lambda)$. Indeed,

$$2\xi L/\pi = -(2L/\pi) \cdot (sq + q^2)/k = -[4L/(n\Lambda)] \cdot [\theta_{inc} + 0.5|\theta_{diff}|] = -[4L/(n\Lambda)] \cdot [\theta_{inc} + (\lambda/\Lambda)] \qquad (39)$$

Let us make an estimation of the un-diffracted intensity in the worst case scenario of the normal incidence of illuminating wave, $\theta_{inc} = 0$:

$$(1-\eta) \approx [4L\lambda/(n\Lambda^2)]^2 \equiv \{[8(n_e - n_o)/n] \cdot (L/\Lambda)^2\}^2 \qquad (40)$$

For example, taking $(L/\Lambda) = 0.2$ (below critical value) and $(n_e - n_o)/n = 0.1$, we get $(1-\eta) \approx 1.0 \cdot 10^{-3}$, or $1/(1-\eta) \approx 1000$. This last number characterizes the value of potential contrast ratio, i.e. intensity ratio for bright state of pixel to that for the dark one. This estimate shows that the walk-off effects will not degrade the performance of the suggested device to a considerable degree.

Another potential source of transmission of incident light into zeroth diffraction order is the deviation of equilibrium structure of a non-monoconstant NLC from its ideal profile Eq. (1). Taking the perturbative approximation for those distortions, Eqs. (21-23), one gets additional coupling coefficients between various diffraction orders. Suppose for definiteness that the light, which illuminates the pixel), is an incoherent combination of right- and left-circularly-polarized waves. Then the most "damaging" is the coefficient $\kappa_0(z)$, which couples those orders. In zeroth approximation the amplitude $R_0(z) = R_0(0) \cos(\kappa_1 z) \equiv R_0(0) \cos(\kappa z)$, and then coupling via $\kappa_0$ results in excitation of additional amplitude $\delta L_0(z)$, incoherent with the incident $L_0(0)$;

$$\delta L_0(z = L) = \int_0^L \kappa_0(z) R_0(z) dz, \qquad (41)$$

Calculations show that the distortions given by Eqs. (21-23) lead to the following expression for $\kappa_0(z)$, and thus to the incoherent addition $\delta L_0(z)$:

$\kappa_0(z) = \kappa\{[(K_3 - K_1)/K_2](2\pi L/\Lambda)^2\}[z(L-z)/(4L^2)]; \qquad \delta L_0(z=L) = 0.043i\{[(K_3 - K_1)/K_2](2\pi L/\Lambda)^2\} R_0(z=0).$

where we used the already assumed condition $\kappa L = \pi/2$. Taking the worst case scenario, when $L/\Lambda = 0.3$, i.e. when thickness is close to critical, for 5CB, and recalling that for thickness close to critical the numerically found amplitude of distortion is about 5 times smaller (for 5CB) than the result of analytic calculation by perturbation theory, Eqs. (21-23), we obtain

$$|\delta L_0(z=L)|^2 = |R_0(z=L)|^2 \cdot (0.034/25) \equiv |R_0(z=L)|^2 /727.$$

We see that the deviation of the director profile from the ideal structure Eq. (1) does not significantly harm the contrast ratio.

**5. Conclusion**

To conclude, in this paper we suggest a new design of an element of LC projection display based on transverse periodicity of azimuthal alignment of nematic liquid crystal. We have calculated the conditions of stability of such LC structure as well as the conditions when this element completely eliminates the zeroth order of the incident light via diffraction. Several important factors that might deteriorate the performance and contrast ratio of such an element were considered in detail. Among essential advantages of the suggested device is the possibility to completely utilize the total power of depolarized illuminating light, as well as rejection of the signal using aperture instead of polarizer. Our estimates make us cautionary optimistic about the prospects of creation of such an element.

**6. Acknowledgments**

Authors would like to express their gratitude to S. T. Wu, M. G. Moharam, R. S. Hakobyan, M. Reznikov and O. Lavrentovich for the discussions of the subject of the paper. The work was partially supported by the Army Research Laboratory Grant # DAAD 190210009.

15## 7. References

[1] Shin-Tson Wu, Deng-Ke Yang, *Reflective Liquid Crystal Displays*, 352 pp., John Wiley & Sons, 2001

[2] Pouchi Yeh and Claire Gu, *Optics of Liquid Crystal Displays*, , John Wiley & Sons, 1999

[3] Tai-Ran Hsu, *MEMS and Microsystems: Design and Manufacture*, McGraw-Hill, 2001

[4] R. Raue, et al., *Phosphor screens in cathode-ray tubes for projection television*, Philips Technical Review, 44, No. 11/12, 1989, p. 335.

[5] M. Bouvier, T. Scharf, *Analysis of nematic-liquid-crystals binary gratings with high spatial frequency*, Optical Engineering, 39, #8, 2129-2137 (2000).

[6] C.J. Yu, J.H. Park, J. Kim, M.S. Jung, S.D. Lee, *Design of binary diffraction gratings of liquid crystals in a linearly graded phase model*, Applied Optics, 43, #9, 1783-1788 (2004).

[7] J.H. Park, C.J. Yu, J. Kim, S.Y. Chung, S.D. Lee, *Concept of a liquid-crystal polarization beamsplitter based on binary phase gratings*, Applied Physics Letters, 83, #10, 1917-1920 (2003).

[8] S.J. Walker, J. Jahns, L. Li, W.M. Mansfield, P. Mulgrew, D.M. Tennant, C.W. Roberts, L.C. West, N.K. Ailawadi, *Design and fabrication of high-efficiency beam splitters and beam deflectors for integrated planar micro-optic systems*, Applied Optics, 32, #14, 2494-2501 (1993).

[9] M.L. Jepsen, H.J. Gerritsen, *Liquid-crystal-filled gratings with high diffraction efficiency*, Optics Letters, 21, #14, 1081-1083 (1996).

[10] B.Ya. Zeldovich, N.V. Tabirian, *Devices for displaying visual information*, Disclosure, School of Optics / CREOL, July 2000.

[11] E. Hazman, V. Kleiner, A. Niv, G. Biener, *Quantized Pancharatnam-Berry phase diffractive optics by use of Space-variant subwavelength grating*; CLEO-QELS 2003, paper # QTuL7.

[12] H. Sarkissian, J.B. Park, N.V. Tabirian, B.Ya. Zeldovich, *Periodically aligned liquid crystal: potential application for projection displays and stability of LC configuration*, Optics in the Southeast 2003 conference program, p. PSE 02.
1515

Figure captions to the paper Periodically Aligned Liquid Crystal: Potential application for projection displays, by H. Sarkissian, N. Tabirian, B. Park, and B. Zeldovich.

Figure 1. Orientation of Nematic Liquid Crystal in a planar cell $0 \leq z \leq L$, with the azimuth $\varphi(x)$ of the director monotonously growing along the transverse coordinate: $\varphi(x) = qx \equiv 2\pi x/\Lambda$. Figure 1a: top view. Figure 1b: side view.

Figure 2. Design of a pixel of a projection display operating on periodically aligned NLC. When no voltage is applied across the cell, light of both polarizations is 100% diffracted into $\pm 1$–st orders and is blocked by the aperture of the projection lens. Application of voltage results in suppression of diffraction and thus in the bright state of the pixel.

Figure 3. Instability of the ideal periodically aligned structure Eq. (1) of NLC arises at the thickness $L < \alpha\Lambda$, where the dimensionless parameter $\alpha$ characterizes the threshold. Figure 3a: monoconstant NLC, for which both analytic and numeric calculations yield $\alpha = 0.5$. Figure 3b: derivative of pervious function better illustrates the sharp onset of instability.

Figure 4. Instability of the periodically aligned structure Eq. (1) of NLC for the non-monoconstant NLC: particular values were taken as for 5CB. Figure 4a: symmetry-breaking perturbation onsets at $L > 0.33\Lambda$. Figure 4b: derivative of pervious function better illustrates the sharp onset of instability.

Figure 5. Onset of Friedericksz transition in periodically aligned liquid crystal: volume average value of $\langle (n_z)^2 \rangle$ versus parameters $L/\Lambda$ and applied voltage $V$. Figure 5a: monoconstant NLC. Figure 5b: graph for the 5CB.

Figure 6. Towards the calculation of the walk-off effects in the diffraction process.

Table 1. Critical thickness, maximum diffraction angle, and Friedericksz transition voltage for a number of NLC's (input data was taken from Ref. [2]).

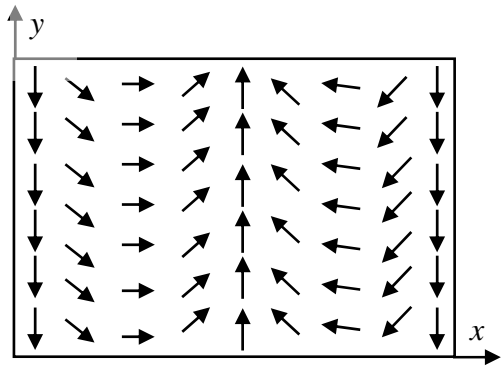
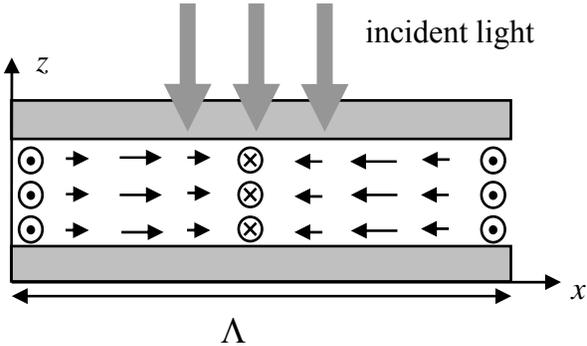

Figure 1a
Figure 1b

Figure 1. Orientation of a Nematic Liquid Crystal in a planar cell $0 \leq z \leq L$, with the azimuth $\varphi(x)$ of the director monotonously growing along the transverse coordinate: $\varphi(x) = qx$. Figure 1a: top view. Figure 1b: side view.

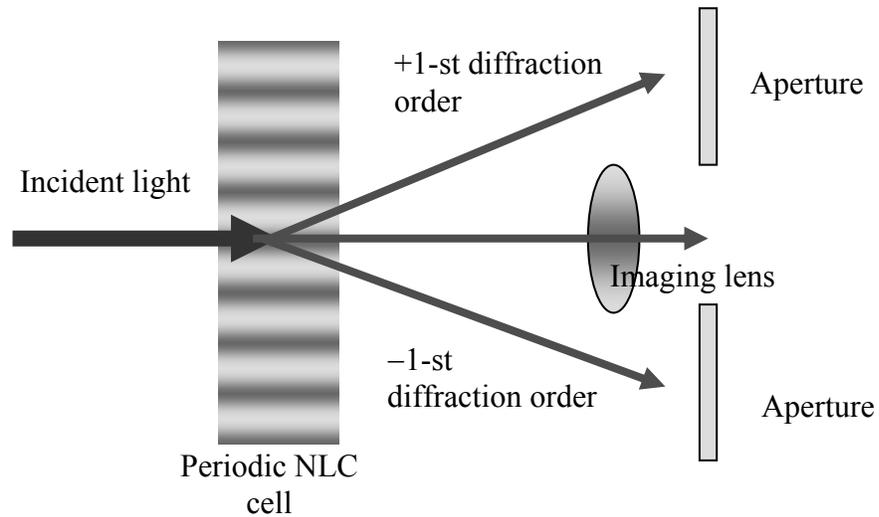

Figure 2. The design of a pixel of a projection display using periodically aligned NLC cell. When no voltage is applied to the cell, 100% diffraction of light of both polarizations into ±1–st orders results in their elimination by an aperture of the projection lens. Application of voltage results in suppression of diffraction and thus in the bright state of the pixel.

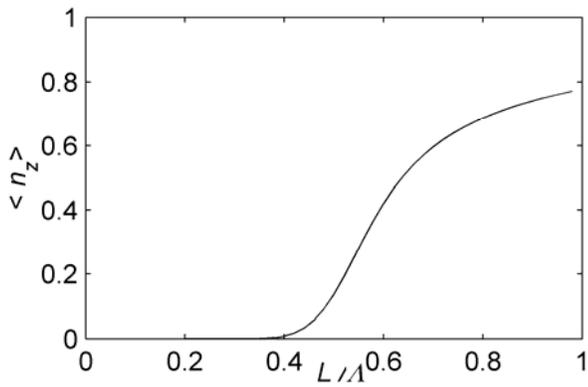 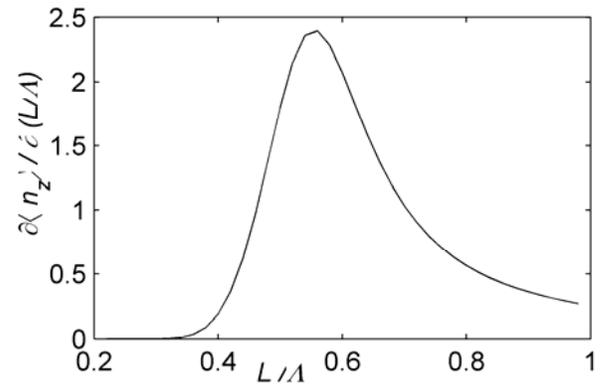

Figure 3a	Figure 3b

Figure 3. Instability of the ideal periodically aligned structure Eq. (1) of NLC arises at the thickness $L < \alpha \Lambda$, where the dimensionless parameter $\alpha$ characterizes the threshold. Figure 3a: monoconstant NLC, for which both analytic and numeric calculations yield $\alpha = 0.5$. Figure 3b: derivative of pervious function better illustrates the sharp onset of instability.

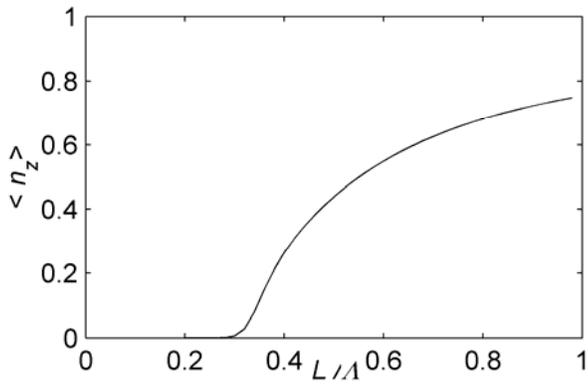 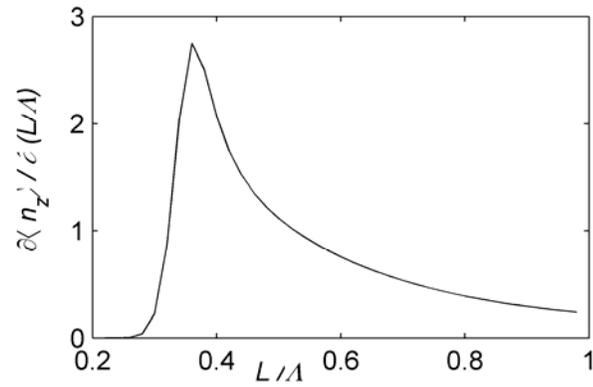

Figure 4a    Figure 4b

Figure 4. Instability of the periodically aligned structure Eq. (1) of NLC for the non-monoconstant NLC: particular values were taken as for 5CB. Figure 4a: symmetry-breaking perturbation onsets at $L > 0.33\Lambda$. Figure 4b: derivative of pervious function better illustrates the sharp onset of instability.

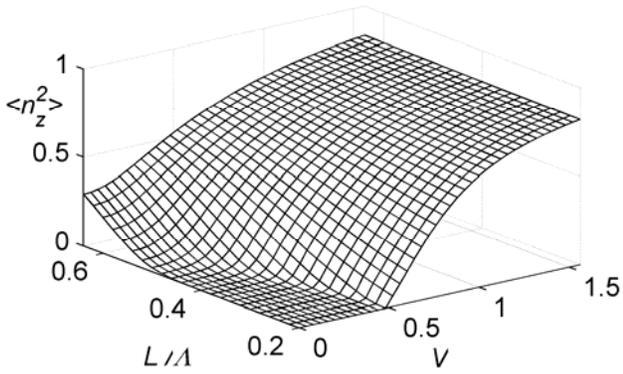 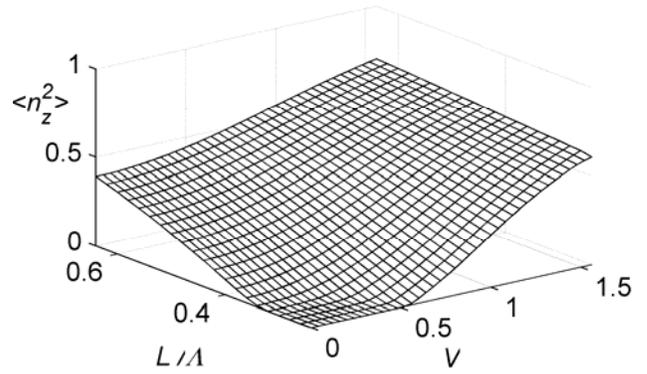

Figure 5a    Figure 5b

Figure 5. Onset of Friedericksz transition in periodically aligned liquid crystal: volume average value of $\langle (n_z)^2 \rangle$ versus parameters $L/\Lambda$ and applied voltage $V$. Figure 5a: monoconstant NLC. Figure 5b: graph for the 5CB.

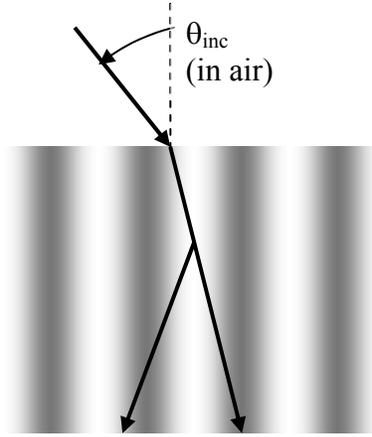

Figure 6.

Figure 6. Towards the calculation of the walk-off effects in the diffraction process.

Table 1. Critical thickness, maximum diffraction angle, and Friedericksz transition voltage for a number of NLC's (input data was taken from Ref. [2]).

| LC type | λ (nm) | Δn | $\varepsilon_{//}$ | $\varepsilon_\perp$ | $K_1$ | $K_2$ | $K_3$ | Friedericksz voltage (Volt) | Critical thickness (Λ) | Diffraction angle (°) |
|---|---|---|---|---|---|---|---|---|---|---|
| | | | | | \multicolumn{3}{c}{$(10^{-12}$ N)} | | | |
| MBBA | 589 | 0.22 | 4.7 | 5.4 | 6.2 | 3.8 | 8.6 | N/A | 0.39 | 20.1 |
| PCH-5 (T=30.3ºC) | 589 | 0.12 | 17 | 5 | 8.5 | 5.1 | 16.2 | 0.88 | 0.33 | 8.9 |
| PCH-5 (T=38.5ºC) | 589 | 0.11 | 17 | 15 | 7.3 | 4.5 | 13.2 | 2.5 | 0.33 | 8.3 |
| PCH-5 (T=46.7ºC) | 589 | 0.1 | 16 | 5.7 | 5.9 | 3.9 | 9.9 | 0.8 | 0.35 | 8 |
| K15 (5CB) | 515 | 0.19 | 20 | 6.7 | 6.4 | 3 | 10 | 0.74 | 0.33 | 14.7 |
| K21 (7CB) | 577 | 0.16 | 16 | 6 | 5.95 | 4 | 6.6 | 0.83 | 0.43 | 15.6 |
| M15 (5OCB) | 589 | 0.19 | 18 | 6.7 | 6.1 | 3.74 | 8.4 | 0.78 | 0.39 | 17.5 |
| E7 | 589 | 0.23 | 20 | 5.1 | 12 | 9 | 19.5 | 0.96 | 0.37 | 19.6 |
| ZLI-1646 | 589 | 0.08 | 11 | 4.6 | 7.7 | 4 | 12.2 | 1.2 | 0.35 | 6.4 |